\documentclass[12pt,preprint]{aastex}

\usepackage{lscape}

\newcommand{\rl}{radio-loud~}

\newcommand{\rqu}{radio-quiet~}

\shortauthors{ P. Grandi}

\begin{document}




\title{Broad Line Radio Galaxies: Jet Contribution to the nuclear X-Ray Continuum}







\author{Paola Grandi}

\affil{Istituto di Astrofisica Spaziale e Fisica Cosmica IASF-Bologna, INAF, Via Gobetti 101, Bologna, I-40129 Italy}

\author{Giorgio G.C. Palumbo}
\affil{Dipartimento di Astronomia, Universit\'a di Bologna, via Ranzani 1, 40127 Bologna, Italy}

\email{grandi@iasfbo.inaf.it,  giorgio.palumbo@unibo.it}










\begin{abstract}

It is shown that, for Broad Line Radio Galaxies the strength of the
non-thermal beamed radiation, when present, is always smaller than the
accretion flow by a factor $< 0.7$ in the 2-10
keV band. The result has been obtained
using the procedure adopted for disentangling the Flat Spectrum  Radio Quasar
3C 273 (Grandi $\&$ Palumbo 2004)  Although this implies a significantly smaller non-thermal flux
in Radio Galaxies when compared to Blazars, the jet component, if present, could be
important at very high energies and thus easily detectable with GLAST.

\end{abstract}

\keywords{galaxies: active galaxies ---- X-rays: radio galaxies; jets}

\section{Introduction}

The \rqu versus \rl dichotomy in Active Galaxies (AGNs) has been confuted
by new deeper radio surveys which have revealed 
a continuous distribution of radio--loudness (White et al. 2000).
Nuclear ejection of matter in the form  of outflows and weak radio jets 
have also been observed in sources generally considered radio quiet AGNs
(Ulvestad $\&$ Wilson 1984a, 1984b, 1989, Ulvestad et al. 2005, Kukula et al. 1995, Thean et al. 2001, Middelberg et al. 2004).
Moreover the well consolidated nuclear correlation between X-ray and radio
emission in powerful \rl sources has been recently extended to a larger sample
of AGNs (Merloni et al. 2003, Maccarone Gallo $\&$ Fender 2003) including
\rqu objects. This  growing evidence in the recent literature has, therfore, put in a critical position 
the long debated idea that Active Galactic Nuclei are divided in two 
sharp classes.
All these results are hinting towards a scenario in which accretion and
ejected flows are closely related in all AGNs implying common physical 
mechanisms. Thus the fundamental  question to address is how 
these mechanisms work under different physical conditions.
In particular, the relation between the launch/quench of the jet and  the disk
accretion regime, is of primary importance.

Broad Line Radio Galaxies (BLRGs) are a key class for addressing this problem.
Unlike blazars, the jet of these sources is not directly pointing towards
the observer. Not completely blinded 
by non-thermal Doppler enhanced radiation, the observer can then attempt to 
look at both the accretion disk and jet. 
Moreover, unlike NLRG/HEGs (Hardcastle et al. 2006), BLRGs are not generally obscured by
large amount of circumnuclear cold gas and, as a consequence, offer 
a direct view of their inner regions.  

Although it is plausible that jets and disks are both present in BLRGs,
results from detailed spectral X-ray studies are not conclusive (Grandi
Malaguti $\&$ Fiocchi 2006; GMF06).
GMF06 analyzed Beppo-SAX  broad band spectra of 35 radio loud
AGNs, 10 of which are BLRGs, and  provide evidence of an anti-correlation
between reflection line equivalent width (EW) and Radio Core Dominance (R)
as expected if the beamed jet contribution is not negligible.
However large errors from poor statistics cast some doubts on the
robustness of the anticorrelation. 

The present paper attempts to disentangle jet and disk in 3
Broad Line Radio Galaxies.
The work was inspired by the 3C 273 results (Grandi $\&$ Palumbo, 2004) in
which, separating  jet from disk physical components, allowed a logical and self 
consistent picture to emerge for all the observed
parameters of this blazar . 
In principle the approach, taken to tackle the problem in 3C 273, can be 
readily applied to other AGNs, provided data taken over a broad band (from 0.1
keV to at least 100 keV) and with sufficiently good statistics are available.

\section{Extending the 3C 273 method to BLRGs}

Analysis of BeppoSAX data of 3C 273 allowed 
to decouple the beamed non-thermal
(jet) and unbeamed thermal (accretion flow) radiation produced 
in the inner region of a blazar.
Previous approaches were phenomenological, i.e  data were treated with ``ad
hoc''models in order to reproduce the observed X-ray spectra.
The new analysis adopted a different starting-point.
Data were fitted, having in mind  a physical picture of the nuclear region. 
Thus, a black body, a cutoffed power law with a reflected component
and an iron line, accounting for an accretion disk, plus a power law,
reproducing a jet was attempted. The experiment, performed on different 
BeppoSAX observations of 3C 273,  was successful and  revealed
a thermal component generally overwhelmed by non-thermal radiation
by a factor of 1.2-3  in the 2-10  keV range
and up to a factor of 7 above 20 keV. 

The successive natural step has been to verify whether the same diagnostic
approach could work for  other radio loud AGNs. 
The best sources suitable for this test are  3C 120,
3C 382 and 3C390.3. Their BeppoSAX data cover a wide energy range 
and their spectra show continuum high energy cutoff and reprocessing features 
simultaneously observed (see GMF06). 

In order to separate thermal (T, i.e. accretion flow) and 
non-thermal (NT, i.e. jet) contributions, detecting features over the continuum is
essential. Indeed the model anchors to the cutoff energy, iron line
equivalent width and reflection hump to estimate the jet contribution.
After assuming that a disk emission is diluted by Doppler enhanced 
radiation.
Furthermore it is implicitly assumed that the disk in BLRGs and Seyfert 1 galaxies
produce similar continuum and reprocessing features (see later in Discussion).  

No ``physical'' hypothesis was put forward for the soft component, 
observed in 3C 120 and 3C 382. Unlike the case of 3C 273, where a disk origin 
of the soft excess (below a few keV) is quite convincing, 
the nature of the soft photons in these sources 
is still uncertain (Zdziarski $\&$ Grandi 2001, Grandi et al. 2001, Ogle et al.
2005). 

From the practical point of view, the following procedure was applied to each
BLRG.
The \textsc{pexrav} in XSPEC v11.1 and a gaussian line
were used to fit the direct continuum and the reprocessed features
of the accretion flow. The inclination angle ($i$)
of the reflecting cold matter was always fixed to $i=18^\circ$ ($\cos\;i =
0.95$).  Although a different inclination angle can not be excluded in these 
radio galaxies, the face-on geometry was chosen to allow for a coherent
comparison  with the previous BeppoSAX survey results, 
not including the jet contribution. On the other hand, given the errors 
associated to the data (Table 5 of GMF06), an inclination angle change of about a factor 2 is included when we explore the range of parameters in Table 1.
The line was assumed cold and narrow 
(energy peak and intrinsic width) in agreement with GMF06 results (see their 
Table 4).  The accretion disk parameters were fixed to the average 
Seyfert 1 values, i.e $\Gamma^T$=1.79, $E_{cut}^T$=166 keV,
$Ref^T=0.75$ and $EW^T=137$ eV 
deduced by a  sample of 7 Seyfert 1 galaxies 
observed by BeppoSAX (see  Table 5 of GMF06) and analyzed by Perola et al. (2002).
The normalization (i.e the flux) was let free to vary.
The jet was modeled with a power law. Spectral slopes ( $\Gamma^{NT}$) of 
1.5, 1.6, 1.7, 1.8 and 1.9 were tested. Each time,  $\Gamma^{NT}$  was fixed
while the normalization (i.e the jet flux) was let free to vary.
As discussed above, the soft component was not investigated. 
A simple ``phenomenological''model was considered: a Raymond-Smith model and a  power
law were used for
3C 120 and 3C 382, respectively.
In the case of 3C 382 when the presence of the jet model was investigated, a broken
power law, rather than a double power law, was tested in order to minimize
the number of free parameters. 
In 3C 120 and 3C390.3, known to be slightly obscured sources, a intrinsic column density (NH)
was added in addition to the Galactic one.

A disk-jet model was considered acceptable (on the base of a F test) if
statistically undistinguished or better than the 
model without a jet component (i.e that adopted in the BeppoSAX survey paper).

Finally, the same test was repeated for each minimum 
and maximum values  of $\Gamma^T$, $E_{cut}^T$,
$Ref^T$ and $EW^T$ defining the  $1\sigma$ distribution spread 
reported in Table 5 of GMF (2006). In summary, 405 disk-jet possible
configurations were considered as shown in Table 1. 

\footnotesize
\begin{deluxetable}{ccccc}
\tablecolumns{5}
\tablewidth{0pc}
\tablecaption{Model Tested$^a$}
\tablehead{
\multicolumn{4}{c}{Thermal/Accretion }  &
\multicolumn{1}{c}{Non-Thermal/Jet } \\
\multicolumn{4}{c}{Model (T)}& \colhead{Model (NT)}\\
\colhead{}              & \colhead{}              & \colhead{}           & \colhead{}       &\colhead{}\\
\colhead{$\Gamma^T$}    & \colhead{$E_{cut}^T$}   & \colhead{Ref$^T$}    & \colhead{EW$^T$} & \colhead{$\Gamma^{NT}$} \\
\colhead{}              & \colhead{(keV)}         & \colhead{}           & \colhead{(eV)} & \colhead{}}
\startdata
1.79 & 166 & 0.75       & 137 [93, 266]& 1.5-1.9\\
1.79 & 166 & 0.45       & 137 [93, 266]& 1.5.1.9\\
1.79 & 166 & 1.05       & 137 [93, 266]& 1.5.1.9\\
1.79 & 93  & 0.75       & 137 [93, 266]& 1.5-1.9\\
1.79 & 93  & 0.45       & 137 [93, 266]& 1.5.1.9\\
1.79 & 93  & 1.05       & 137 [93, 266]& 1.5.1.9\\
1.79 & 239 & 0.75       & 137 [93, 266]& 1.5.1.9\\
1.79 & 239 & 0.45       & 137 [93, 266]& 1.5.1.9\\
1.79 & 239  &1.05       & 137 [93, 266]& 1.5.1.9\\
1.65 & 166 & 0.75       & 137 [93, 266]& 1.5-1.9 \\
1.65 & 166 & 0.45       & 137 [93, 266]& 1.5.1.9\\
1.65 & 166 & 1.05       & 137 [93, 266]& 1.5.1.9\\
1.65 & 93  & 0.75       & 137 [93, 266]& 1.5-1.9\\
1.65 & 93  & 0.45       & 137 [93, 266]& 1.5.1.9\\
1.65 & 93  & 1.05       & 137 [93, 266]& 1.5.1.9\\
1.65 & 239 & 0.75       & 137 [93, 266]& 1.5.1.9\\
1.65 & 239 & 0.45       & 137 [93, 266]& 1.5.1.9\\
1.65 & 239  &1.05       & 137 [93, 266]& 1.5.1.9\\
1.93 & 166 & 0.75       & 137 [93, 266]& 1.5-1.9 \\
1.93 & 166 & 0.45       & 137 [93, 266]& 1.5.1.9\\
1.93 & 166 & 1.05       & 137 [93, 266]& 1.5.1.9\\
1.93 & 93  & 0.75       & 137 [93, 266]& 1.5-1.9\\
1.93 & 93  & 0.45       & 137 [93, 266]& 1.5.1.9\\
1.93 & 93  & 1.05       & 137 [93, 266]& 1.5.1.9\\
1.93 & 239 & 0.75       & 137 [93, 266]& 1.5.1.9\\
1.93 & 239 & 0.45       & 137 [93, 266]& 1.5.1.9\\
1.93 & 239  &1.05       & 137 [93, 266]& 1.5.1.9\\
\enddata
\tablenotetext{(a)}{Each row represent 15 different disk-jet configurations, 
obtained considering 3 Fe EW (the average value [minimum,maximum]) and 5 spectral slopes of the jet ($\Gamma^T$=1.5, 1.6, 1.7, 1.8, 1.9)}
\end{deluxetable}
\normalsize

\section{Results}

In spite of the wide grid of models tested, the successful trials are limited.
These are shown in Table 2 with 
the ratio between thermal and non-thermal fluxes (T/NT) between 2-10 keV.
The addition of a NT component always gives an acceptable fit. Although, the
degrees of freedom of the model without the non-thermal component
are larger, the $\chi^2$ values are not significantly better than the disk-jet
combination. In some cases, disk+jet fit gives  $\chi^2/dof$ even smaller.
Fits requiring 
$\Gamma^T \sim \Gamma^{NT}$ are not reported in Table 2. 
No useful information is contained in these trials. They simply split the
continuum photons in two very similar power laws, thus imitating the ``phenomenological'' model.

\begin{deluxetable}{ccccccccc}
\tabletypesize{\scriptsize}
\tablecolumns{9}
\tablewidth{0pc}
\tablecaption{Fit Results}
\tablehead{
\multicolumn{1}{c}{Source }&
\multicolumn{4}{c}{Thermal/Accretion}  &
\multicolumn{1}{c}{Non-Thermal/Jet} &
\multicolumn{1}{c}{Flux ratio}&
\multicolumn{1}{c}{$\chi^2$/d.o.f.} &
\multicolumn{1}{c}{$\chi^2$/d.o.f.}\\
\colhead{}&
\multicolumn{4}{c}{Model (T)}&
\colhead{Model (NT)}&
\colhead{$NT/T$}&
\colhead{With NT}&
\colhead{Without NT} \\
\colhead{}&
\colhead{$\Gamma^T$}&
\colhead{$E_{cut}^T$ (keV)}   &
\colhead{Ref$^T$}    &
\colhead{EW (eV)} &
\colhead{$\Gamma^{NT}$}    &
\colhead{(2$-10$ keV)}&
\colhead{} &\colhead{}}
\startdata
3C 120  &1.79 &  93  & 0.45 & 90                & 1.4  &   $<0.09$              & 132/133 & 129/130 \\
        &1.79 &  93  & 0.45 & 90                & 1.5  &   $<0.1$               & 133/133 & 129/130 \\
        &1.79 &  93  & 0.45 & 90                & 1.6  &   $<0.2$               & 133/133 & 129/130 \\
        &1.93 &  93  & 0.75 & 136               & 1.5  &    0.3$^{+0.1}_{-0.1}$ & 135/133 & 129/130 \\
        &1.93 &  93  & 0.75 & 136               & 1.6  &    0.5$^{+0.3}_{-0.3}$ & 135/133 & 129/130 \\ 
        &1.93 &  93  & 1.05 & 136                & 1.4  &      0.2$^{+0.1}_{-0.1}$   & 135/133 & 129/130 \\
        &1.93 &  93  & 1.05 & 136                & 1.5  &      0.2$^{+0.1}_{-0.1}$   & 133/133 & 129/130 \\
        &1.93 &  93  & 1.05 & 136               & 1.6  &       0.4$^{+0.1}_{-0.1}$   & 131/133$^{\star}$ & 129/130\\ 
        &1.93 &  93  & 1.05 & 136               & 1.7  &       0.7$^{+0.1}_{-0.1}$   & 134/133 & 129/130 \\
&&&&&&&&\\
3C 390  &1.79 &  166 & 0.75 & 136               & 1.4  &  0.3$\pm0.1$     & 125/120 & 124/117\\
        &1.79 &  166 & 0.75 & 136               & 1.5  &  0.4$\pm0.2$     & 128/120 & 124/117 \\
        &1.79 &  166 & 1.05 & 136               & 1.4  &  0.2$\pm0.1$       & 122/120$^{\star}$ & 124/117 \\
        &1.79 &  166 & 1.05 & 136               & 1.5  &  0.3$\pm0.2$       & 123/120 & 124/117 \\
        &1.79 &  166 & 1.05 & 136               & 1.6  &  0.4$\pm0.3$       & 128/120 & 124/117 \\ 
        &1.79 &  93  & 0.75 & 136               & 1.4  &  0.4$\pm0.1$     & 128/120 & 124/117 \\
        &1.79 &  166 & 0.75 & 90               & 1.4  & 0.3$\pm0.1$      & 131/120 & 124/117 \\
&&&&&&&&\\
3C 382      &1.79 &          93  &      0.45   & 90  &   $^a1.5$  &     0.13$\pm0.04$ &149/149       &  138/146 \\  
            &1.79 &          93  &      0.45   & 90  &   $^a1.6$  &     0.5$\pm0.2$   &146/149           &  138/146 \\
            &1.93&           93  &      0.75   & 90  &   $^a1.5$  &     0.3$\pm0.1$   &149/149           &  138/146 \\
            & 1.93&          93  &      0.75   & 90  &   $^a1.6$  &     0.5 $\pm0.1$  &142/149          &   138/146\\  
            & 1.93&          93  &      0.75   & 90  &   $^a1.7$  &     1.0 $\pm0.3$  &145/149           &   138/146 \\
            & 1.93&          93  &      1.05   & 90  &   $^a1.7$  &     0.8 $\pm0.3$  &143/149          &   138/146 \\
            & 1.93&         166  &      0.75   & 90  &   $^a1.7$  &     0.7$\pm0.2$   &147/149          &  138/146 \\
            & 1.93&          93  &      0.75   & 90  &   $^a1.7$  &     1.1$\pm0.3$   &149/149            &  138/146 \\
 &1.8$^{+0.1}_{-0.6}$& 85$^{+142}_{-25}$&0.5$^{+0.5}_{-0.3}$&$^c38^{+37}_{-36}$& $^b1.5^{+0.9}_{-0.4}$&0.2$^{+0.5}_{-0.2}$&
135/145$^{\star}$ &  138/146 \\

\enddata
\tablenotetext{(a)}{ The soft slopes ($\Gamma_{\rm soft}$) of the broken power law range between 2.1-2.7. The energy  
breaks ($E_{\rm break}$) are between  2.2 and 3.0 keV}
\tablenotetext{(b)}{ $\Gamma_{\rm soft}= 2.7^{+0.5}_{-0.6}$, $E_{\rm break}=
  2.8^{+2.0}_{-0.8}$}
\tablenotetext{(c)}{ The Fe equivalent width is calculated considering only
  the thermal X-ray continuum }
\tablenotetext{(\star)}{Disk-jet configurations with the smallest $\chi^2$ values  }
\end{deluxetable}
\normalsize
{\it 3C 120} -- The 3C 120 data allow 9 different jet-disk configurations. All
require a low energy cut-off ($E_{cut}^T=93$ keV), quite steep 
thermal power law ($\Gamma_{T}= 1.79-1.93$) and two possible values of 
the iron line equivalent width ( EW$^T$=90-136 eV). 
When the reflection and the iron lines are weak only upper limits ($NT/T<0.4$)
of the jet strength can be obtained (see 1-4 rows in Table 2). 
On the contrary, when stronger iron lines and reflection humps are assumed, the
jet contribution to the X-ray continuum is well constrained. It is of the
order of  $\sim 20-40\%$, depending of the slope of the non-thermal power
law. The strength of the non-thermal component increases with the steepening of
$\Gamma^{NT}$ . Note that the spectral feature which 
more strongly constrains the jet-disk
configuration is the high energy cutoff. 

\noindent
{\it 3C 390.3} -- In this source, the Doppler enhanced non-thermal radiation
requires a flat spectral slope ($\Gamma_{NT}= 1.4-1.6$) and a $NT/T$ flux
ratio in the 2 -10 keV ranging from 0.1-0.7 (taking into account the flux
ratio uncertainties). Also in this case 
the jet contribution to the X-ray continuum can not be larger than  $40\%$.

\noindent
{\it 3C 382} -- The soft excess of this source is rather complex.
However it can be simplified by adopting  a pure power law. 
As discussed  when the jet component is also taken into account a broken power
law is introduced. The successful models are reported  in Table 2. 
The most relevant result here is that, in spite of the jet powerful $NT/T=0.1-1.1$, 
the iron line is forced to its minimum permitted value (EW$^T$=90 eV). 

Finally, it should be noted that, unlike the other 2 
BLRGs,  for 3C 382 one can estimate the jet power fixing ``a priori'' all 
the spectral parameters (last row in Table 2). Obviously, the uncertainties 
are large and include all the previous disk-jet configurations.
The only exception is the Fe equivalent width, which, being smaller than 90 eV, confirms
the extreme weakness of the reprocessed feature in this source.
\section{Discussion}

On the bases of the BeppoSAX results one cannot exclude that a non-thermal component is present 
in BLRGs. Actually the fits obtained for 3C 120 3C390.3 and 3C 382 
show that the data are consistent with a combination of a
thermal component (in a first approximation associated  with an accretion disk) and 
a non-thermal component to be associated with beamed radiation (i.e a jet).
However the jet contribution is not large and
the accretion flow emission in the 2-10 keV range ($NT/T< 0.7$) is always dominant.
In 3C 382 a very powerful jet ($NT/T>1$) is statistically permitted by the model.
However its contribution to the X-ray continuum becomes
($NT/T \le 0.7$), matching the other BLRG values, as soon as the disk-jet
spectral parameters are let free to vary. 
3C 382 provides further information. Its Fe feature is weak 
even  if a jet is included to fit the data.
This implies that the production of features by a disk and/or a dusty torus 
is really inefficient, suggesting that a dilution of the X-ray continuum 
due to a Doppler enhanced radiation is not the main cause of the Fe line shrinking.

The successful trials in Table 2 display generally
hard non-thermal spectral slopes ($\Gamma^{NT} \sim 1.5-1.7$). Thus the jet of
these BLRGs could be  in the
Inverse-Compton regime. This sounds promising. Indeed in the Spectral
Energy Distributions (SED) of these radio galaxies (Figure 1) there is a bump
in the mm-infrared band which could hide the low energy synchrotron peak.
Although part of the infrared radiation is probably a mix of 
thermal (from cold/dusty matter) and non-thermal radiation, 
this peak associated to the jet X-ray spectrum undoubtedly recalls 
the  synchrotron-Inverse Compton SED of a blazar.

\begin{figure}
\epsscale{1.0}
\plotone{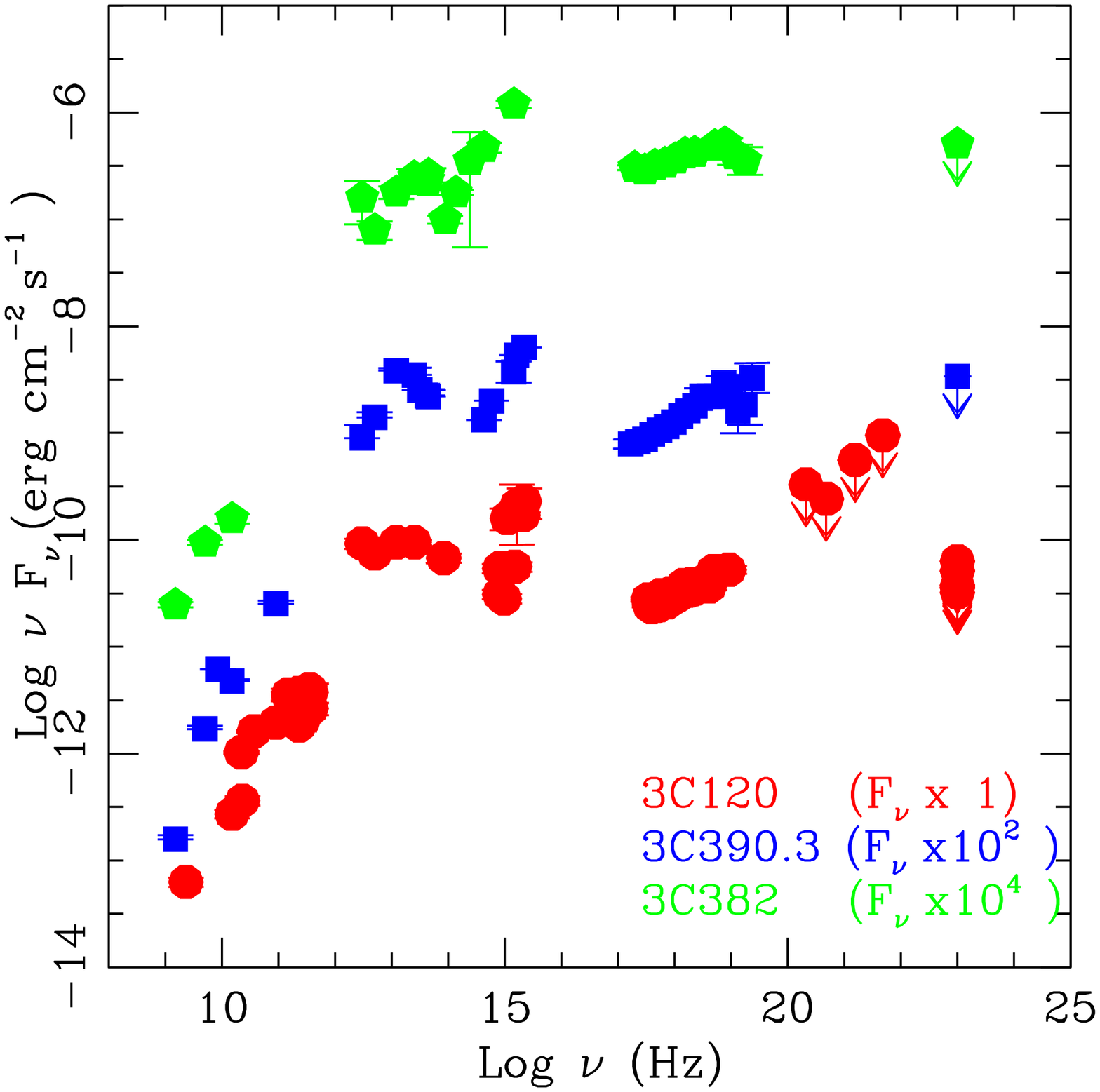}
\caption{Spectral Energy Distribution of 3C 120,  3C390.3 and 3C 382.
With the exception of the BeppoSAX data (this paper), the SED points
are from literature: Bloom et 1994, Chiaberge et al.2002,  Clavel et al. 2000, 
Hardcastle et al. 1999, Lilly $\&$ Longair 1985, Knapp et al.1991, Mc Alary et al. 1979,Maisack et al. 1997, Maraschi et al.  1991, 
Morganti et al. 1997, Robson et al. 2001, Rudnick et al. 1996,Schachter $\&$ Elvis 1994, 
 Steppe et al. 1990, 1991, Tadhunter et al.1986, Von Montigny et al. 1995, Zirbel et al. 1993.Optical--UV data are corrected for Galactic/Intrinsic reddening  (Cardelli et al. 1989) assuming the 
BeppoSAX column density $N_H$. }
\label{sed}
\end{figure}

Although the observed X-ray photons in the 2-10 keV band are produced at most
in an accretion flow, the jet, if present, is expected 
to prevail over the thermal radiation above 100-200 keV.
Indeed, at high energies, the accretion flow 
power law drops and  the jet should emerge.
The successive immediate question to address is: are the jets in BLRGs
 bright enough to be directly observable at high energies?
In order to verify it,  we extrapolated the jet-disk models
up to GeV energies.
For each BLRG, we considered as the most probable description  of the 
data, the model in Table 2  with the smallest $\chi^2$ value (indicated with
$\star$).  It appeared immediately evident that the jet 
spectrum has to bend before reaching the GeV region to respect the 
EGRET upper limit constraint (see Figure 2). 
This is not surprising. The 3C 273 beamed component, which has been 
actually detected by both the Comptel (Von Montigny et al. 1996) and EGRET
(von Montigny et al. 1993) instruments on-board of the GRO satellite, shows exactly that spectral shape.
Taking advantage from this, the Compton peak and the GeV emission of each 
radio galaxy was reconstructed following the 3C 273 curvature.
In details, the non-thermal 100 keV flux density of 3C 273 
was divided by each BLRG and the ratio used to rescale the 3C 273 spectrum to 
the jet emission of the three Radio Galaxies.
The results are shown in Figure 2, where the MeV-GeV extrapolated fluxes
of each radio galaxy are drawn as open circles. 
Although BLRG are less luminous by about a factor 10 than 3C 273, their 
non-thermal radiation will become easily detectable for telescopes of  the near future.
In Figure 2 the GLAST integral sensitivity curve\footnote{$http://www-glast.slac.stanford.edu/software/IS/glast\_lat\_performance.htm$} 
rescaled to 55 days of observation is shown (solid blue line).
Even considering a very pessimistic reduction by a factor 2 of the 
on-flight GLAST performance (dotted blue line), we conclude that 
less than 8 weeks should be  sufficient for GLAST to reveal radio galaxies with
fluxes of order $F_{E>100 MeV}= 4\times 10^{-8}$ phots $cm^{-2} s^{-1} MeV^{-1}$.

\begin{figure}
\epsscale{1.0}
\plottwo{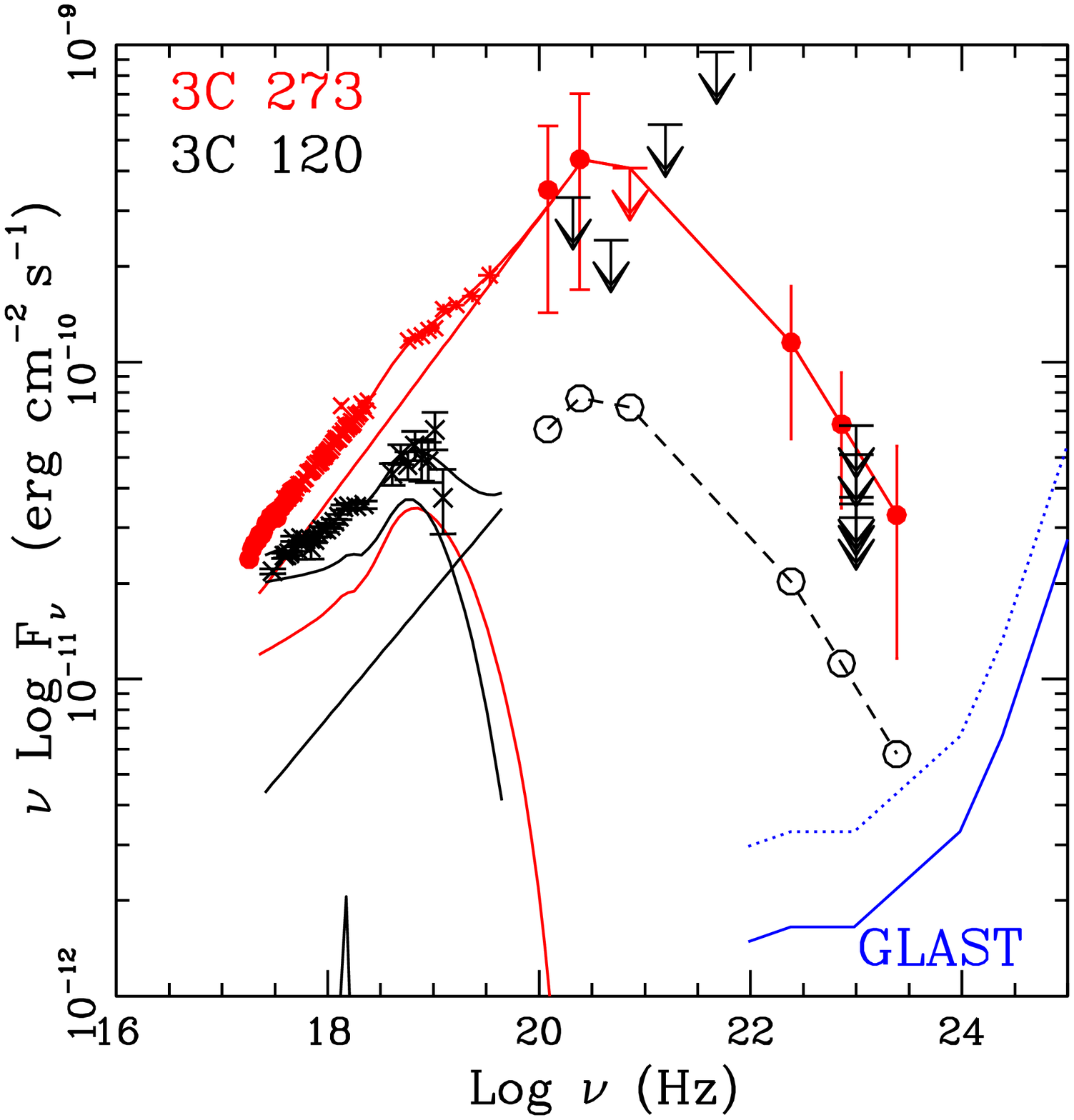}{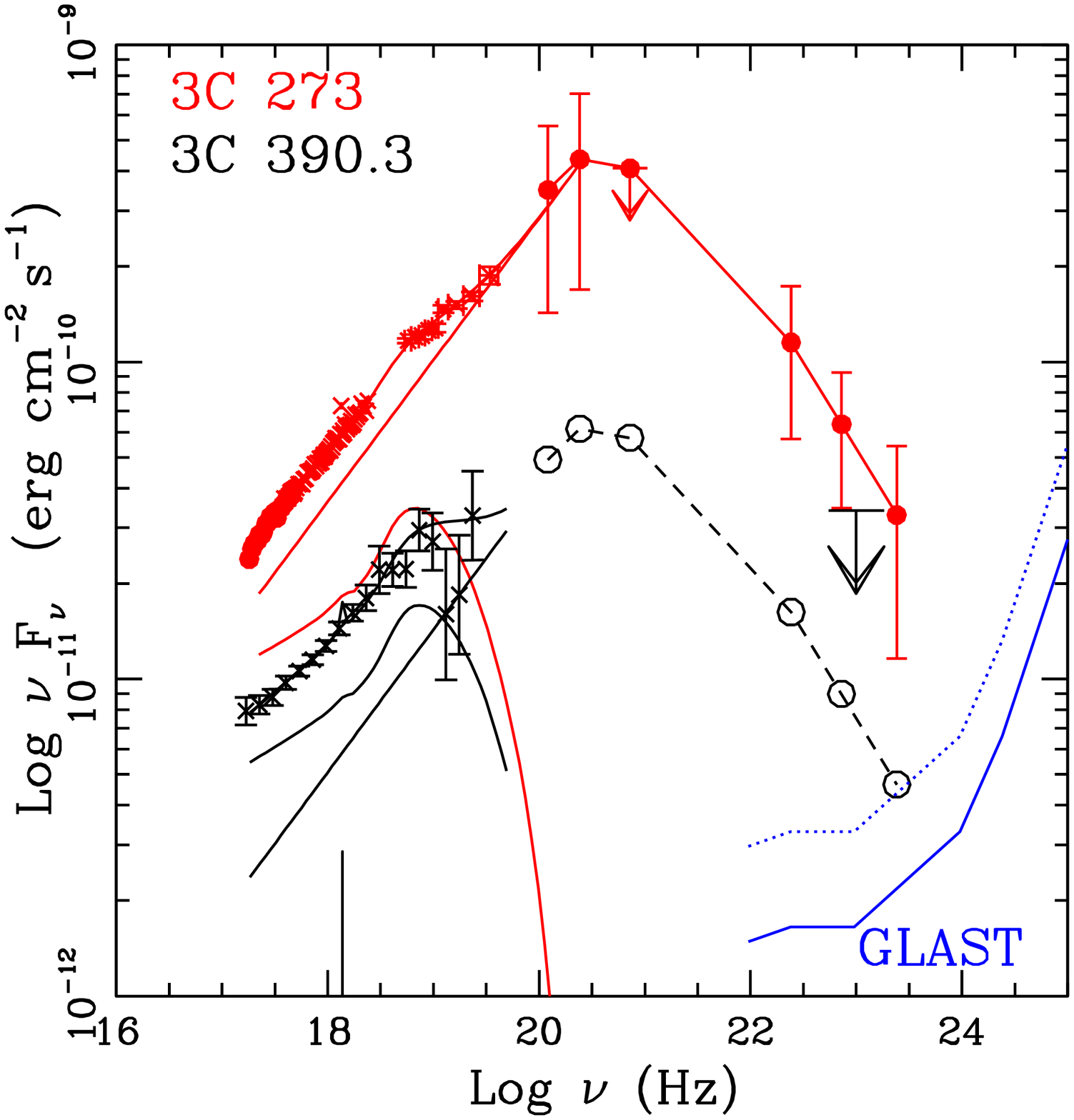}
\epsscale{0.45}
\plotone{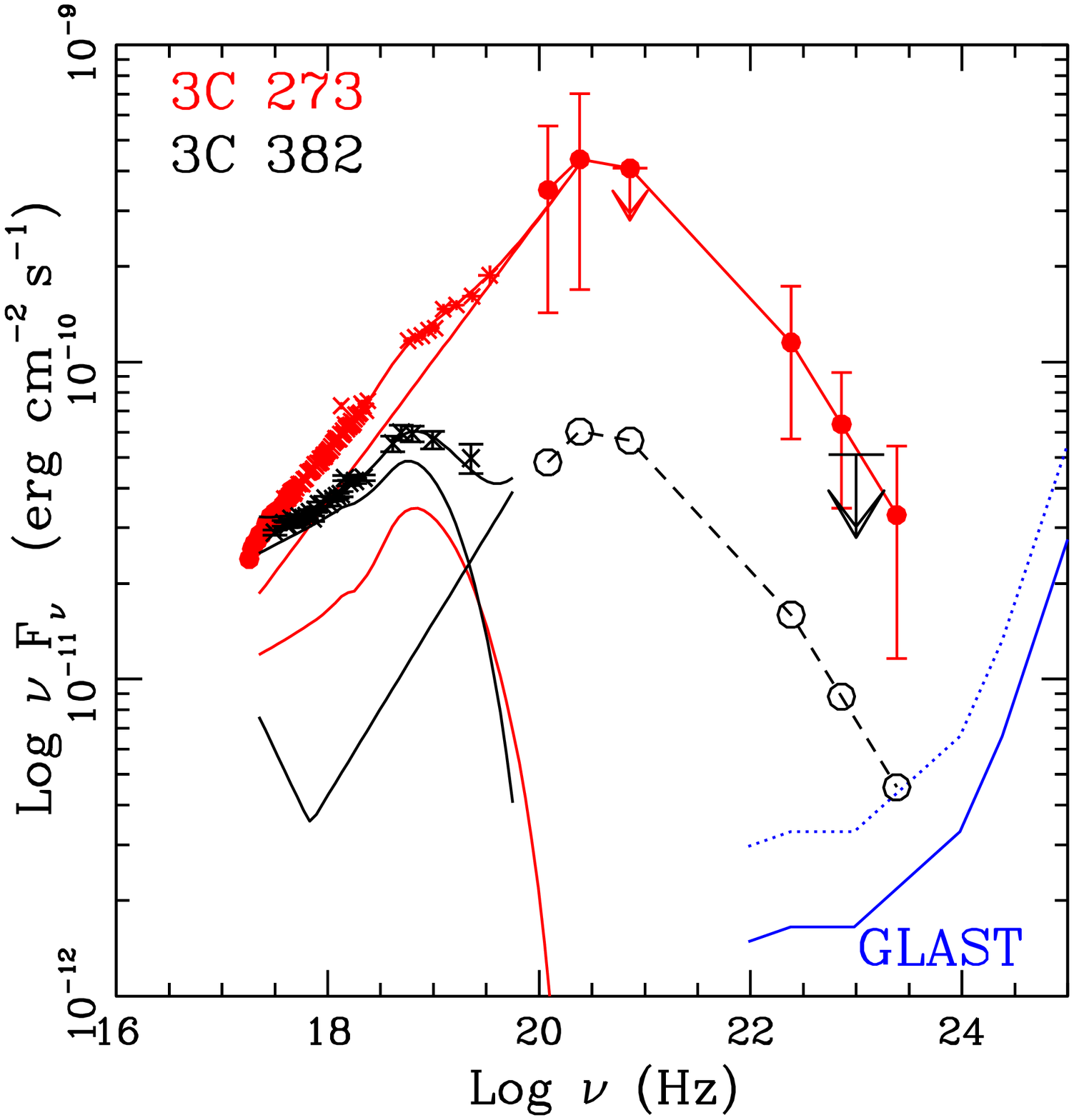}
\epsscale{0.5}
\caption{BeppoSAX BLRG (black) and 3C 273 (red) data (EGRET data from Turler
  et al. 1999) fitted with a
combination of accretion flow and jet models (solid lines).  
Open circles represent 
the jet extrapolation to high energies, assuming a 3C 273 spectral
shape. Black dotted lines are point interpolations. 3C 120 (left panel), 3C390.3  (right panel) and 3C 382 (lower panel) will
be all detectable by GLAST. Solid blue line represents the 55 days 
GLAST integral sensitivity curve. Even considering a less optimistic
sensitivity curve, rescaled by a factor
2 (dotted blue line), BLRG jets should be detected by GLAST}
\label{sed}
\end{figure}

\section{Conclusion}

A new interpretation of the BLRGs spectra  observed by BeppoSAX is discussed.
The method successfully applied to the Blazar 3C 273 
(Grandi $\&$ Palumbo 2004) to disentangle  the jet and  disk components has
been used. Following this approach for  3C 120, 3C390.3 and 3C 382, we conclude that:
\noindent
\begin{itemize}
\item{Beamed non--thermal X--ray continuum does not significantly contaminate  the nuclear 
emission of these sources. The jet, if present, does not  contribute more than $45\%$.}
\item{The iron line is weak even taking into account the presence of a non-thermal beamed 
radiation,  at least in 3C 382. If confirmed by other BLRG observations
(e.g. the Suzaku satellite), 
other physical reasons rather than the jet presence, have to be invoked.
For example, assuming that the disk is thin but optically thick, 
less prominent iron lines could be explained within a light bending model as proposed 
by Miniutti $\&$ Fabian 2004. Alternatively one can assume that in radio-galaxies the accreting
gas is characterized by low radiative efficiency (at least in the inner regions) and 
explain the weak line as due to small solid angle subtended by the cold matter (i.e external 
disk region and/or  torus) to the primary X-ray source (GMF06 and references therein).} 
\item{The SEDs of powerful BLRGs likely hide a jet with a spectral shape
very similar to that of 3C 273. The infrared bump of radio galaxies 
recalls the synchrotron peak of blazars and the quite hard spectral slope of their 
simulated jet, a non-thermal inverse Compton emission. If this is the case, these BLRGs
are expected to be less luminous than 3C 273 by about a factor 10 in the MeV-GeV regions.
In spite of that, 3C 120, 3C390.3 and 3C 382 are very promising targets for the GLAST mission}
\end{itemize}
  
\begin{acknowledgments}
This research has made use of the NASA/IPAC Extragalactic Database (NED) which is operated by 
the Jet Propulsion Laboratory, California Institute of Technology,
under contract with the National Aeronautics and Space Administration.
We thank Elisabetta Cavazzuti and Carlotta Pittori 
for supporting us in our GeV exploration. 

\end{acknowledgments}

\clearpage



\end{document}